\documentclass[journal]{IEEEtran}
\usepackage[flushleft]{threeparttable}
\usepackage{booktabs}
\ifCLASSINFOpdf
\else
\fi
\usepackage{cite}
\usepackage{amsmath,amssymb,amsfonts}
\usepackage{algorithmic}
\usepackage{graphicx}
\usepackage{textcomp}
\usepackage[flushleft]{threeparttable}
\usepackage{caption}
\usepackage{subcaption}
\usepackage{booktabs}
\usepackage{multirow}
\usepackage{caption}
\usepackage{footnote}
\usepackage{subcaption}
\usepackage{acronym}
\usepackage{array}
\usepackage[draft]{hyperref}
\usepackage[font=small,skip=0pt]{subcaption}
\usepackage[font=small,skip=0pt]{caption}
\usepackage{tikz}
\usepackage[export]{adjustbox}

\begin{document}
\title{Improving Heart Rate Estimation on Consumer Grade Wrist-Worn
Device Using\\Post-Calibration Approach}
\author{Tanut Choksatchawathi, Puntawat Ponglertnapakorn, Apiwat Ditthapron, Pitshaporn Leelaarporn, \\Thayakorn Wisutthisen, Maytus Piriyajitakonkij, and Theerawit Wilaiprasitporn  \IEEEmembership{Member, IEEE} \vspace{-2em}
\thanks{This work was supported by Robotics AI and intelligent Solution Project, PTT Public Company Limited, Thailand Science Research and Innovation (SRI62W1501), and Thailand Research Fund and Office of the Higher Education Commission (MRG6180028).}
\thanks{Tanut Choksatchawathi, Puntawat Ponglertnapakorn, Pitshaporn Leelaarporn, Maytus Piriyajitakonkij and Theerawit Wilaiprasitporn are with Bio-inspired Robotics and Neural Engineering Lab, School of Information Science and Technology, Vidyasirimedhi Institute of Science \& Technology, Rayong, Thailand {\tt\small(corresponding author: theerawit.w at vistec.ac.th)}}
\thanks{Apiwat Ditthapron is with Computer Department, Worcester Polytechnic Institute, Worcester, MA, USA.}
\thanks{Thayakorn Wisutthisen is with School of Information Technology, King Mongkut's University of Technology Thonburi, Bangkok, Thailand}
}


\maketitle

\begin{abstract}
The technological advancement in wireless health monitoring through the direct contact of the skin allows the development of light-weight wrist-worn wearable devices to be equipped with different sensors such as photoplethysmography (PPG) sensors. However, the motion artifact (MA) is possible to occur during daily activities. In this study, we attempted to perform a post-calibration of the heart rate (HR) estimation during the three possible states of average daily activity (resting, \textcolor{red}{laying down}, and intense treadmill activity states) in 29 participants (130 minutes/person) on four popular wearable devices: Fitbit Charge HR, Apple Watch Series 4, TicWatch Pro, and Empatica E4. In comparison to the standard measurement (HR$_\text{ECG}$), HR provided by Fitbit Charge HR (HR$_\text{Fitbit}$) yielded the highest error of $3.26 \pm 0.34$ bpm in resting, $2.33 \pm 0.23$ bpm in \textcolor{red}{laying down}, $9.53 \pm 1.47$ bpm in intense treadmill activity states, and $5.02 \pm 0.64$ bpm in all states combined among the four chosen devices. Following our improving HR estimation model with rolling windows as feature (HR$_\text{R}$), the mean absolute error (MAE) was significantly reduced by $33.44\%$ in resting, $15.88\%$ in \textcolor{red}{laying down}, $9.55\%$ in intense treadmill activity states, and  $18.73\%$ in all states combined. This demonstrates the feasibility of our proposed methods in order to correct and provide HR monitoring post-calibrated with high accuracy, raising further awareness of individual fitness in the daily application.
\end{abstract}

 

\IEEEpeerreviewmaketitle

\section{Introduction}
\IEEEPARstart{H}{eart} rate (HR) is one of the most convenient measurements that reflects the performance of the cardiovascular system and the overall health condition \cite{palatini2011role}. The abnormal escalation of HR indicates a possibility of any failures in the physique, but can often be detected and diagnosed prior to the emergence of the symptoms traditionally by monitoring the HR using an electrocardiogram (ECG) \cite{palatini2011role, reil2011heart, parak2015evaluation, nystoriak2018cardiovascular}. Daily activities can be categorized into a number of states with different physical activity levels (PAL), in which the extracted physical information can be used to forecast the possibility of having cardiovascular disease (CVD) \cite{mukhopadhyay2015wearable}. For instance, a study screening for metabolic dysfunctions by Fernandes \textit{et al.} \cite{fernandes2013resting} demonstrated a correlation between HR during the resting period and the risk of having CVD. Dyslipidemia and high blood glucose were found to be correlated with high HR during passive activity in adolescents. Similarly, a low HR during recovery state, a duration after an intense exercise, has been shown to indicate a possibility of coronary artery disease \cite{ghaffari2011abnormal}. Recently, many successful diagnostic processes of cardiac arrhythmia such as atrial fibrillation have adapted smart wearable devices to aid HR monitoring and further health deterioration prevention \cite{dorr2019watch}. 

With the advanced technology, the concept of wearable devices has increased the mobility of the users, expanding the utilization of the apparatus into the daily life application beyond the confinement of the clinical establishments \cite{tedesco2019accuracy, sawangjai2019consumer, lakhan2019consumer, autthasan2019single}. 
Many wearable ECG devices provide HR estimation with validated accuracy, however, with a requirement of contact to the chest area using a chest strap which reduces mobility, causes inconvenience, and limits the application of the devices \cite{spierer2015validation}. First described in 1937 \cite{hertzman1937observations}, Photoplethysmography (PPG) was introduced to estimate HR (HR$_\text{PPG}$) with the benefit of mobility and the ease of integration to ubiquitous devices \cite{asada2003mobile}. There are two types of PPG sensors: transmission and reflection \cite{ram2011novel, joseph2014photoplethysmogram}. A transmission type PPG sensor measures the light intensity on the other side of the light source. The reflection PPG sensor generally contains at least one LED and one photo-diode on the same side of the device as a base to detect a relatively small change in reflected light affected by the blood flow. The PPG signal reflects only a small pulsatile portion; only 0.1\% of total signal amplitude\cite{ram2011novel}. Subsequently, various noise components, such as ambient light, the electromagnetic coupling from the other sensors, and motion artifacts (MA), generated by the changes of the distance in the gap between a device and the skin, often disturb the PPG signals \cite{joseph2014photoplethysmogram}. Devices with PPG were previously reported to underestimate the HR$_\text{PPG}$ during an intense activity as movements occurred, preventing the detection of the peak-to-peak interval used to count the heartbeat per minute \cite{bai2018comparative, wallen2016accuracy}. Due to the higher probability of MA while wearing a wearable device, HR estimation by PPG may not yield as accurate readings, and therefore, needs further improvement. These techniques, requiring movement-related information, include an adaptive algorithm for minimum noise generation \cite{yousefi2013motion}, sparse signal reconstruction \cite{zhang2014troika}, and multi-channel spectral matrix decomposition \cite{xiong2016spectral}.

The effort to ameliorate the frameworks for HR monitoring and MA removal and reduction has led to the collection of datasets stored in various databases \cite{biswas2019heart}. In 2015, the IEEE Signal Processing Cup (SPC) database initiated the assembly of the state-of-the-art HR recordings from PPG sensors during different sets of exercises. Several frameworks, consisting of combinations of HR$_\text{PPG}$ estimation techniques, are being evaluated. The TROIKA framework, using sparse signal reconstruction, was tested on a dataset recorded from 12 subjects and was shown to perform with high estimation accuracy \cite{zhang2014troika}. Other frameworks, e.g., WPFV \cite{temko2017accurate} using Wiener filter and phase vocoder, particle filter \cite{nathan2017particle}, and CorNET \cite{biswas2019cornet} using deep learning algorithms, were explored to alleviate the MA. Although the endeavor to correct HR using raw accelerometer and gyroscope has been challenged by previous studies \cite{lee2018motion,pamula2018system,albadawi2018heart,culbert2016motion}, none has exploited the motion sensing elements such as physical activity, step count, and motion pattern, as featured in the improving HR estimation model. These previous works have been anchoring and assuming the existence of possible practical information from the raw data reported by the PPG sensors as well as equipped inertial measurement unit (IMU) motion sensors. However, commercialized devices generally do not grant permission for the alteration of the HR estimation algorithm and the removal of the prominent MA constituents.

The aforementioned studies unquestionably confirmed the benefit of using the accelerometer and/or gyroscope to curtail the undesired MA and correct the PPG signals. However, the PPG signals, logged as a set of time-series data, can be relatively massive, leading to the prominent concern over the storage capacity of any wearable device \cite{winfree2017modeling, henriksen2018using}. Hence, most of the consumer grade wearable devices often record only the HR$_\text{PPG}$. Therefore, we presented a post-calibration method using the information obtained from these wrist-worn devices as well as additional personal information from the users. We proposed two regression models, with and without the rolling window, to improve the HR$_\text{PPG}$ using the built-in sensing constituents as the main features as an attempt to achieve the most accurate HR estimation. Instead of executing artifact removal on the PPG signal, a robust post-calibration method was developed in which it can be directly applied to the derived HR$_\text{PPG}$ with MA residue. Feature selection was performed on the selected devices by testing the linear correlation and mutual information of each feature to the HR$_\text{ECG}$. Without the rolling window, six machine learning (ML) algorithms were formulated and trained with tuned hyperparameters. The rolling regression model has been developed to improve the ultimate estimated HR, abbreviated as HR$_\text{R}$, from the devices using the recorded movement data, attained directly from the devices.

Our contributions of this study can be summarized as:

\begin{enumerate}
\item In addition to the state-of-the-art datasets \cite{zhang2014troika} in which only the recorded HR$_\text{PPG}$ during the intense treadmill activity state were monitored, we demonstrate the validations of HR measurement in three states including resting, \textcolor{red}{laying down}, and intense treadmill activity states performed by the four wrist-worn devices: Fitbit Charge HR \cite{Fitbit}, Apple Watch Series 4 \cite{Applewatch}, TicWatch Pro 
\cite{TicWatch}, and E4 \cite{EmpaticaE4}. 
   
\item In order to estimate HR, a novel approach of feature selection from a list of candidates, containing the extracted HR$_\text{PPG}$ (the raw HR provided) from the selected wrist-worn device, PAL, step count, gender, Pittsburgh sleep quality index (PSQI), and Body Mass Index (BMI), was introduced. The input features were chosen using a univariate linear regression with a null hypothesis testing, while the non-linear relationship was processed using the mutual information derived from the ECG (HR$_\text{ECG}$). The physical activity levels (PALs), step count, and rolling windows were subsequently exploited as the main features of the testbed to improve the HR$_\text{PPG}$.

\item The proposed process of post-calibration included further evaluation of the temporal information as features from the rolling windows. The rolling regression was also verified to enhance our proposed methods of HR estimation (HR$_\text{R}$) tested on our dataset of 29 participants (130 mins/participant). All results from the selected models (HR$_\text{ML}$ from ML models, HR$_\text{SF}$ from sensor fusion, and HR$_\text{R}$ from the rolling regression) were compared to demonstrate the most suitable post-calibration methods for HR estimation.






\end{enumerate}

The remainder of this study is organized into seven sections. The background of different candidate features and the ML algorithms are described in Section \ref{Background}. Our set of experiments in Section \ref{Methods} were divided into two main parts: physical data collection and data processing. The results of all the analyses are reported in Sections \ref{result:validation} and \ref{result:calibration}. The importance of the accuracy assessment and the improvement of HR estimation during the post-calibration process is further elaborated in Section \ref{Discussion}, followed by Section \ref{Conclusion} to conclude the proposed post-calibration methods of HR estimation and its accomplishments.

\section{Background}
\label{Background}
The accuracy of HR measurement is required for assessing the heart and the overall health status. Due to the uniqueness of our physical differences, we first determined a list of measurable candidate features exhibiting the potential to improve HR for each individual. A description of each possible feature is provided in this section, followed by the ML algorithms for regression problem: support vector regression (SVR), random forest (RF), Gaussian process (GP), artificial neural network (ANN), Logistic regression (LR) and k-Nearest Neighbors regression (kNN).

\begin{table*}[]
\centering
\caption{Comparison between technical specification of each wearable devices. \small{(* Raw data not available, ** No application in this study.)}}
\label{table:device-compare}
\begin{tabular}{lcccccc}
\hline
\hline
Device & Sensor & LED & Photodiode & 3-axis Accelerometer & Gyroscope & Price [\$]\\
\hline
Apple Watch Series 4 \cite{Applewatch} & ECG* + PPG & 4 (Green) 2 (Infrared) & 8 & Yes & Yes & 399.00\\
Empatica E4 \cite{EmpaticaE4} & PPG & 2 (Green) 2 (Red) & 2 & Yes & No & 1,690.00\\
Fitbit Charge HR \cite{Fitbit} & PPG & 2 (Green) & 1 &  Yes** & No & 150.00\\
TicWatch Pro \cite{TicWatch} & PPG & 2 (Green) & 1 & Yes & Yes & 249.99\\
Polar H10 \cite{PolarH10} & ECG & - & - & Yes & No & 89.95\\
Biosignalsplux kit \cite{Biosignalsplux} & ECG & - & - & Yes & No & 1,349.93
  \\ \hline \hline

\end{tabular}
\vspace{-5mm}
\end{table*}

\subsection{Wearable devices}
Three consumer grade wrist-worn wearable devices (Fitbit Charge HR \cite{Fitbit}, Apple Watch Series 4 \cite{TicWatch}, and TicWatch Pro \cite{TicWatch}), and one medical grade device class 2a (E4) \cite{EmpaticaE4} wristband available in the market were systemically validated against each other using our proposed protocols for the first time. The technical information for each device is exhibited in Table \ref{table:device-compare}. Fitbit (Fitbit Inc., San Francisco, CA, USA) has been listed as the best seller of the consumer grade wrist-worn wearable products and has been used in validation studies twice as many times as the other brands and 10 times more often than the other 131 brands used in clinical trials \cite{winfree2017modeling, henriksen2018using}. Apple (Apple Inc., Cupertino, CA, USA) dominates over a majority share in the technology sector. The PPG sensor in Apple Watch Series 4 contains more LEDs compared to the other products used to maximize the detection of a pulse wave which may contribute to the higher HR estimation accuracy. The HR$_\text{PPG}$ from TicWatch Pro (Mobvoi Information Technology Company Limited, Beijing, China) can be acquired by their software. TicWatch Pro operates with Wear OS from Google Inc., in which it supports Google service. The device has an energy saving mode, allowing the battery to last for up to 30 days within a single charge. E4 wristband from Empatica (Empatica Inc., Milano, Italy) publicly discloses one of the two algorithms that are parts of the HR calculation. The device also provides Electrodermal Activity (EDA) as well as Infrared to measure the skin temperature. Additionally, two ECG devices were used as the standard measurement of HR stemmed from the ECG (HR$_\text{ECG})$: Biosignalsplux kit (PLUX Wireless Biosignals S.A., Lisbon, Portugal) \cite{Biosignalsplux} and Polar H10 (Polar Electro, Kempele, Finland) \cite{PolarH10}. 

\subsection{Candidate features}
\label{bg:features}
\subsubsection{Physical activity level (PAL)}
\label{section:PAL}
PAL is an estimation of the required physical activity in a day, which can be calculated by dividing the total energy expenditure (TEE) by basal metabolic rate (BMR) \cite{joint2004human}. BMR refers to the energy expenditure at a standard condition of resting. PALs from wrist-worn devices are assessed by the equipped accelerometer \cite{migueles2017accelerometer}. Its filtered and processed signals of the movement are used to compute the activity count per minute (cpm). The accelerometer-based PAL is computed from the change of the body movement in each axis of accelerometer and its interval \cite{hildebrand2014age}.

The current wearable technology enhances the fitness tracking and the measurement of PAL, allowing the classification and monitoring of the fitness intensity level and daily expenditure energy level in users of various ages and environmental conditions \cite{miller2010estimating, winfree2017modeling, awais2018physical}. Fitbit Charge HR, Apple Watch Series 4, and TicWatch Pro promptly report PALs, calculated from the signals detected by the furnished accelerometers using their restricted algorithms, whereas E4 only grants raw accelerometer count, in which an external method is required to compute PAL from the provided data. There are numerous PAL estimation methods depending on the locations of the devices in the physical activity research field \cite{migueles2017accelerometer}. The methods used in this study were based on the implementation of the four methods proposed by Freedson \textit{et al.} \cite{pamty2005calibration}, Troiano \textit{et al.} \cite{troiano2008physical}, and Crouter \textit{et al.} \cite{crouter2015estimating}, which are four of the most adopted PAL estimation methods included as a comparative experiment of physical activity \cite{awais2018physical}. The methods from Freedson \textit{et al.} \cite{pamty2005calibration} and Troiano \textit{et al.} \cite{troiano2008physical} were designed for hip placement. However, recent work by Knaier \textit{et al.} \cite{knaier2019validation} adopted the method from Troiano \textit{et al.} to perform PAL estimation with the accelerometer on the wrist in comparison to the hip, demonstrating that the algorithms can be adapted for both hip- and wrist-worn. Distinctively, only the two methods from Crouter \textit{et al.} \cite{crouter2015estimating}, using both vector magnitude (VM) and the vertical axis (VA), were originally designed for the PAL measuring on the wrist. These four PAL estimation methods differ in the threshold or the cut-point of the PAL. The cut-points for the four intensities of PAL (sedentary (SED), light physical activity (LPA), moderate physical activity (MPA), and vigorous physical activity (VPA)) were used following the original paper employing ActiGraph accelerometer by Freedson \textit{et al.} \cite{pamty2005calibration}. The four PALs are summarized in Table \ref{table:cutpoint}.

\begin{table}[t]\setlength\tabcolsep{2pt}
\centering
\caption{Physical activity level (PAL) cut-points in count per minute (cpm).}
\label{table:cutpoint}
\begin{tabular}{cccccc}
\hline
\hline
PAL threshold & Vector & SED & LPA & MPA & VPA\\
\hline
Crouter \textit{et al.} & VA &$\leq35$ & $36-360$ & $361-1129$ & $\geq1130$\\
Crouter \textit{et al.} & VM &$\leq100$ & $101-609$ & $610-1809$ & $\geq1810$\\
Freedson \textit{et al.} & VA &$\leq99$ & $100-759$ & $760-5724$ & $5725-9498$\\
Troiano \textit{et al.} & VA &$\leq100$ & $101-2019$ & $2020-5998$ & $\geq5998$
  \\ \hline \hline
\end{tabular}
\vspace{-5mm}
\end{table}

\subsubsection{Step count}

Walking and running are two of the most common physical activities in daily life that consume energy \cite{wilkin2012energy}. In 2009, a study reported a positive relationship between the HR recovery level and the number of steps \cite{lubans2009relationship}. Almost all wrist-worn wearable devices are incorporated with the step counting feature as the main measurement to evaluate the daily activity performance of the user, especially in the fitness trackers. All devices that were used in this study, with the exception of the ECG sensor from Biosignalsplux kit, are equipped with the pedometer to record the steps along with either HR$_\text{ECG}$ or HR$_\text{PPG}$ continuously and simultaneously. Serving a different function from the previously mentioned cpm which can be determined from PAL, the steps per minute were calculated at each HR estimation data point from the aggregated step counting.

\subsubsection{Personal Health information}
Previous studies have reported the correlations between the level of cardiorespiratory fitness and gender \cite{lubans2009relationship, silvetti2001heart}. The measurements of HR in male individuals were found to be significantly higher in female individuals \cite{silvetti2001heart}. In addition to the previous features that contained the body movement information aforementioned in this study, gender was included as background information of each participant along with Body Mass Index (BMI) and Pittsburgh sleep quality index (PSQI) in the candidate feature list. BMI, one of the most measured personal physical information, was reported to be associated with pNN50, a portion of beat-to-beat difference that lasts longer than 50 milliseconds (ms), and the root mean square of successive differences (RMSSD) components of HR variability (HRV) \cite{koenig2014body}. Moreover, there are reports of the correlation between HRV and sleep quality \cite{wei2011subjective,fernandes2013resting} by applying PSQI \cite{buysse1989pittsburgh}. PSQI is a measurement of the sleep quality based on the self-reported sleep evaluation in the form of a questionnaire with the score ranges from 0 to 21. A PSQI that is lower than 5 indicates adequate sleep, whereas a larger score indicates a poor sleep quality. Age was excluded from the candidate feature list as no significant correlation between age and HR has been reported \cite{ryan1994gender}. Furthermore, the age range among the participants was not sufficient in this study.

\subsection{Machine learning algorithms}
\label{sec:ML}
\subsubsection{Support vector regression (SVR)}
\label{sec:R}
As an extension of the support vector machine (SVM), which was introduced to solve a classification problem \cite{cortes1995support}, SVR was derived for a regression problem \cite{drucker1997support}. The SVR approaches the generalized model by minimizing the generalized error bound while building a hyperdimensional function that deviates to the most $\epsilon$ from the training points. To maintain the balance of the generalization and the complexity of the model, a capacity constant ($C$) was introduced as a hyperparameter that needs to be carefully tuned, together with the $\epsilon$. The mathematical detail of this algorithm can be found in a library for SVM \cite{chang2007library} for simplicity in this Section.

\begin{equation}
\label{eq:3}
 \begin{array}{l}
 K_{\text{Poly}}(x_i,x_j) = (\gamma x_i x_j^\top + 1)^d\\
K_{\text{RBF}}(x_i,x_j) = exp(\frac{-\gamma||x_i-x_j ||^2}{2\sigma^2}) = exp(-\gamma||x_i-x_j ||^2)
\end{array}
\end{equation}

In the regression problem, the label $y_i$ can be predicted from the evidence, $x_i,x_j$, which includes the HR$_\text{PPG}$ and features from the wearable devices. The kernel ($K(x_i)$ ) acts as a feature transformation, mapping data points ($x_i,x_j$) to separable spaces in a higher feature space. Polynomial kernel ($K_{\text{Poly}}$) and radial basis function ($K_{\text{RBF}}$) were adopted as in \eqref{eq:3}. The kernels differ from each other in time complexity and parameters that require tuning. The $K_{\text{Poly}}$ has a tunable parameter $d$ that increases the complexity of the kernel, whereas $K_{\text{RBF}}$ has $\sigma$ that controls the generalization of the training data distribution. In all kernels, $\gamma$ is a core parameter that controls the influence of each training instance to the overall model. $\sigma$ in $K_{\text{RBF}}$ is a constant parameter which can be tuned together in the form of $\gamma$.





\subsubsection{Random forest (RF)}
\label{sec:RF}
RF is an ensemble learning method for decision tree (DT) \cite{liaw2002classification}. DT has a set of nodes, articulating hierarchically. One specific feature of DT includes its threshold, making a decision in which a region of a certain sample should be in. Each node creates two divided cuboid subspaces for its input feature. Therefore, the averaged target labels of the training data in each subspace, dividing by the deepest node of each branch, are considered a predicted value. In RF, multiple DT are built, each with a subset of features and training data randomly chosen with a replacement. Subsequently, the average of the results from all DT in RF is reported as a final regression value.

\subsubsection{Gaussian Process (GP)}
\label{sec:GP}
GP is a nonparametric learning method that uses a Bayesian approach to solve a regression problem \cite{rasmussen2006gaussian}. The model is defined by the mean function, often defined as zero, and a covariance kernel ($K$), based on a prior assumption of data. The GP is robust to noise since it is a distribution of multiple random functions. $K$ is a multivariate Gaussian distribution with a zero mean. At inference, GP uses all training data to provide a confidential interval of predicted value, based on the prior belief. The radial basis function kernel ($K_{\text{RBF}}$), as in \eqref{eq:3}, is used as the covariance kernel of GP. In GP, $\sigma$ in $K_{\text{RBF}}$ is considered as a length scale which scales each feature and can be learned from the data.

\subsubsection{Artificial neural network (ANN)}
\label{sec:ANN}
In recent years, ANN has become a well-known machine learning model in both classification and regression tasks of various domains \cite{acharya2003classification}. It outperforms the state-of-the-art traditional ML algorithms using a deep ANN, given that sufficient training data are provided. In this study, the depth of ANN was limited to a shallow network (3-5 hidden layers) to avoid the overfitting problem and generalized the model to work on cross-subject prediction. The fully connected layer architecture, which has hidden layers connecting from the input layer to the output layer with a variety of node's sizes in each layer, was applied.

\subsubsection{Logistic Regression (LR)}
\label{sec:LR}
LR is a statistical model that applies Sigmoid function to the linear combination of the candidate features and sets a threshold for dividing the data into a specific region, working well in binary classification \cite{hosmerdw}. When utilized with a regression task, the algorithm uses the output from Sigmoid function to predict a value without a threshold.

\subsubsection{k-Nearest neighbor regression (kNN)}
\label{sec:kNN}
Commonly known for the classification task, kNN is an approximate algorithm of the Bayesian classifier \cite{altman1992introduction}. The algorithm estimates the unconditional density function of the training data using the kernel method making calculation on k-nearest samples, prior to the prediction of the sample class. Regarding the regression task, the prediction is made by averaging the target labels of k-nearest training samples.

\begin{figure}[b]
\includegraphics[width=85mm]{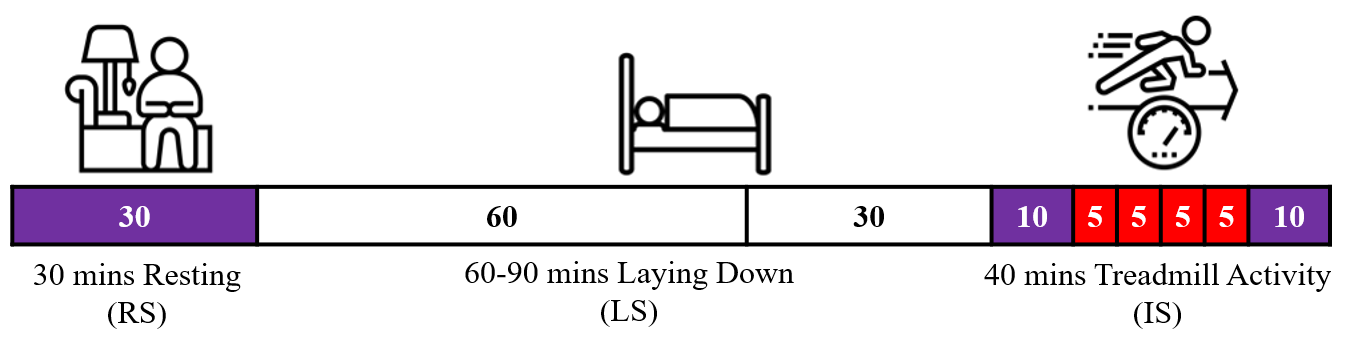}
  \caption{Timeline of the experimental protocol. HR$_\text{ECG}$, HR$_\text{PPG}$, PAL, accelerometer data, and step count were collected during the three activity states (resting for 30 minutes, \textcolor{red}{laying down} for 60-90 minutes, and intense treadmill activities for 40 minutes).}
  \label{fig:experimental}
  \vspace{-6mm}
\end{figure}

\section{Methods}
\label{Methods}
The experimental protocol, as shown in Figure \ref{fig:experimental}, and the data recording are first introduced, followed by the descriptive information of the participants and the statistical analysis. 

\subsection{Data collection}
\label{sec:StandardMeasurement}

The physical activity assessed in this study consists of three states: resting, \textcolor{red}{laying down}, and intense treadmill activity states. Two ECG monitoring devices were employed to collect HR$_\text{ECG}$ as a standard measurement. During the states with low energy expenditure, i.e., resting and \textcolor{red}{laying down} states, a single-lead ECG research grade device from the Biosignalsplux wearable platform is assembled. The ECG signals were wirelessly acquired through OpenSignals software. Although the Biosignalsplux kit has not been validated for clinical purposes, to our best knowledge, BITalino, which is a previously released platform from the same company, had been validated \cite{guerreiro2014performance} and has been used as a standard measurement \cite{tassignon2018continuous}. Despite supported with a high resolution of 16-bit, which can accommodate the sampling frequency up to 3,000 Hz, the Biosignalsplux kit is perhaps not designed for dynamic motion. Therefore, Polar H10, which provides interference-free ECG measurement with a chest strap HR sensor, was used as an alternative during the intense treadmill activity state, involving walking and running in a fixed amount of time interval, and the HR$_\text{ECG}$ obtained was applied as a standard measurement. Polar H7, the predecessor of Polar H10 in the same product line, had been validated previously \cite{plews2017comparison, gillinov2017variable} and was used as a standard measurement in this study \cite{eather2019efficacy}.
\subsection{Experimental Protocol}
The experimental design was \textcolor{red}{investigated} to imitate daily activities. The activities in resting, \textcolor{red}{laying down}, and intense treadmill activity states were devised accordingly, as shown in Figure \ref{fig:experimental}. Before beginning the experiments, skin preparation was performed to improve the quality of the signal acquisition \cite{kligfield2007recommendations}. Skin preparation gel and conductive gel were applied locally on the single-lead ECG placement. Informed consent was received from all participants following the Helsinki Declaration of 1975 (as revised in 2000), which was approved by the internal review board of Rayong Hospital, Thailand.

The recording in the resting state commenced with the participants relaxing in a living room environment while watching a collection of animated videos with a content rating of no violence or nudity present and simple language as the means to stabilize the mental activity that might affect the HR. This follows the reports on the direct relationship between a lower arousal rate and the media content with the recommended age in Parental Guidance (PG) \cite{fleming2001effects}. ``Peppa pig", a children's series which is recommended for PG 3+, was selected to simulate the resting state. The task in the resting state was continued for 30 minutes while the data were simultaneously collected as follows: ECG signal from Biosignalsplux, accelerometer signals from E4, PAL from the three consumer grade wearable devices, and HR$_\text{PPG}$ from Fitbit Charge HR, Apple Watch Series 4, TicWatch Pro, and E4.

After the resting state, the participants were asked to  \textcolor{red}{lay down} in a \textcolor{red}{bedroom-like} environment. \textcolor{red}{
The recording began 5 minutes after the participants started to relax and lay down. The time interval of the experiment were arranged for 60 to 90 minutes.} The data were recorded with the same devices as in the resting state.

A set of physical activities in the intense treadmill activity state was then proceeded after the \textcolor{red}{laying down} state. The participants were instructed to rest for 10 minutes and thereafter stroll on the treadmill at three different speeds (walking: 2 kilometers per hour (kmph), brisk walking: 5 kmph, and jogging: 8 kmph). The duration of the physical cooling down period (walking: 2 kmph) after the main activities lasted for 5 minutes. The speed was slightly increased until reaching the specific speed of each portion. At the end of the set, the participants were instructed to rest for 10 minutes to allow for HR recovery. HR$_\text{ECG}$, HR$_\text{PPG}$, PAL, accelerometer data, and step count were recorded throughout the entire set. Biosignalsplux kit was exchanged with Polar H10 prior to the recording for HR$_\text{ECG}$ as Polar H10 is designed for usage during movement while the Biosignalsplux kit is designed for static activity. The recording protocol in other devices remained the same as in resting and \textcolor{red}{laying down} states.
\subsection{Participant} Healthy 29 individuals, age ranging from 15-33 years old, were recruited (Males: 17, Females: 12) to participate in this study. All participants completed the experimental protocol. The descriptive characteristics of the participants are summarized in Table \ref{table_characteristic}. \textcolor{red}{The participants were required to complete a demographic information form, PSQI, and written informed consent. The annual health checkup was checked to ensure that the participants have no CVD in his/her medical history.}

\begin{table}[bt]
\caption{Descriptive characteristics of 29 participants.}
\label{table_characteristic}
\begin{center}
\begin{tabular}{lr}
\hline\hline
 & Mean (SE)\\
\hline
Age (year) & 24.62 (3.92)\\
Height (cm) & 162.25 (19.01)\\
Weight (kg) & 61.17 (13.96)\\
PSQI Score & 5.79 (2.92)\\
\hline\hline
\end{tabular}
\end{center}
\vspace{-5mm}
\end{table}

\subsection{Statistical analysis}
The HR$_\text{PPG}$ in all three states were collected using three consumer grade devices and one medical grade device. \textcolor{red}{Some data sampling rate was dropped in some devices due to the inconsistency of HR estimation. To establish a proper analysis, we considered only the data from the participants with a consistent sampling rate in all four devices.} The standard measurement from the Biosignalsplux kit was recorded \textcolor{red}{as} raw ECG signaling. A bandpass filter of 15-20 Hz was applied to the ECG signal to remove the unrelated components in the signal. The only remaining R-peak of the QRS complex is the strongest component in the ECG that indicates the heartbeat\cite{parak2011ecg}. With MNE package \cite{gramfort2013meg}, R-R or inter-beat interval was located and transformed to HR$_\text{ECG}$. Instead of smoothing noisy HR$_\text{ECG}$, occurring from the false detection of R-R interval with a moving average, a low-pass filter of 0.05 was applied to remove a sudden change (high frequency) in the HR$_\text{ECG}$. \textcolor{red}{Regarding discrepancy in HR estimation algorithms, the HR$_\text{ECG}$  was delayed for 10 seconds to match the lagged signal between HR$_\text{ECG}$ and HR$_\text{PPG}$ in all devices. On the other hand, Polar H10 promptly provides HR$_\text{ECG}$.} 

The HR$_\text{PPG}$ from all wrist-worn wearable devices were compared with the baseline standard measurements (HR$_\text{ECG}$) from both Biosignalsplux kit and Polar H10. The mean absolute error (MAE) with a standard error of mean (SE) was individually reported for each device in every state. The value was then compared with the HR$_\text{PPG}$ from the other wrist-worn devices. The comparison between each wrist-worn device was reported, entailing the repeated measures of ANOVA, which required the same group of participants. A mean difference from MAE of each pair was computed with the indication of the significant p-value (p$<$0.05). \textcolor{red}{Only a device with the most significant error was used in the next experiment to improve HR in this study.}

\section{Result and Analysis: Heart rate validation} \label{result:validation}
\begin{figure*}[bt]
\centering
\includegraphics[width=0.96\textwidth]{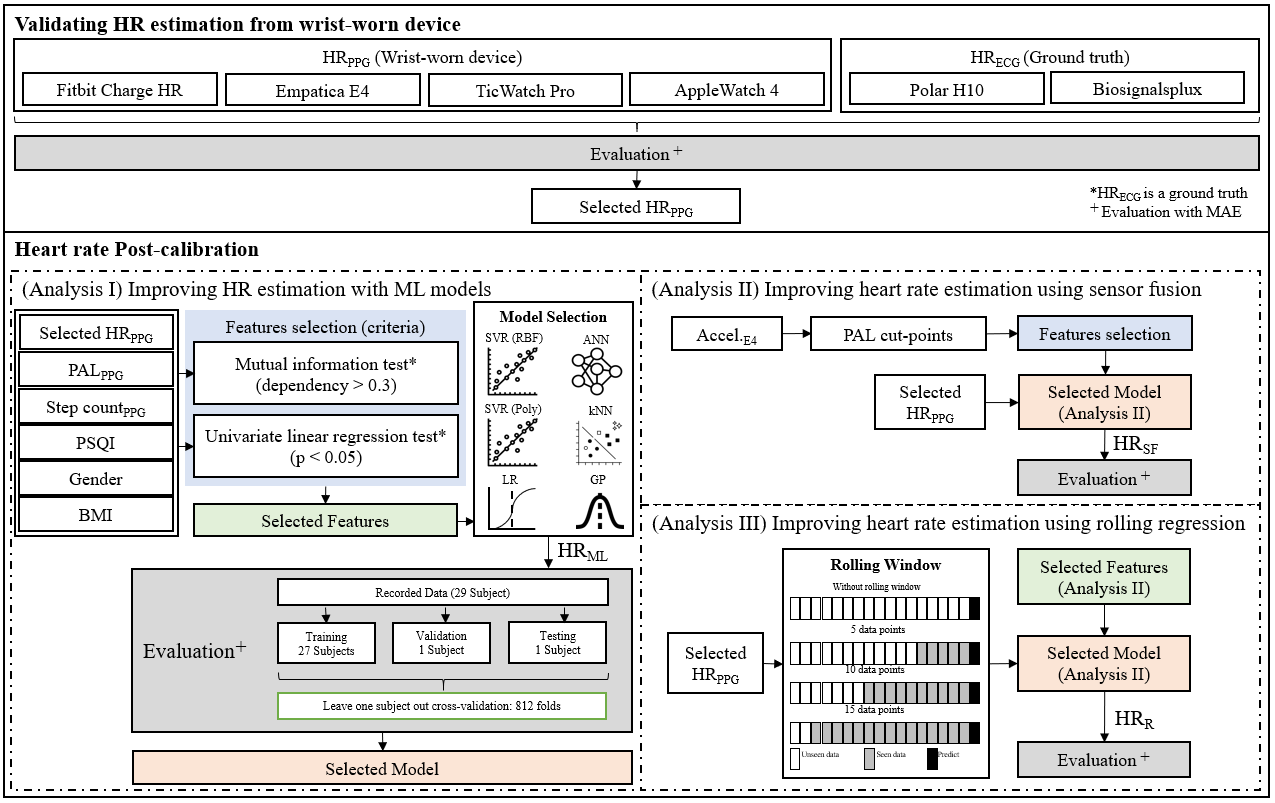}
\caption{Overview diagram of our results and analyses.}
\label{fig:feature_analysis}
\vspace{-0mm}
\end{figure*}

The results of the performed validation of HR$_\text{PPG}$ provided by each wrist-worn device were presented in this section. \textcolor{red}{The MAEs of the baseline measurement HR$_\text{ECG}$ and the four wrist-worn devices were compared, as exhibited in \textit{Validating HR estimation from the wrist-worn device} of Figure \ref{fig:feature_analysis}, to select the device for post-calibration analyses.}
\subsection{Data points collection}
The collected dataset contains 218,356 data points of HR$_\text{PPG}$ from all 29 participants recorded with four wrist-worn wearable devices in resting, \textcolor{red}{laying down}, and intense treadmill activity state. As mentioned previously, only the data from the participants with consistency in the sampling of HR$_\text{PPG}$ measured by all devices were used. The characteristics of the data are shown in Table \ref{table:datasets}. Each device operates with a different sampling rate, depending on the device specification, which varies from 1 Hz to 0.07 Hz. \textbf{Note:} E4 has a built-in algorithm that restrains the output of HR$_\text{PPG}$ if the quality of the observed R-R interval is not detectable.
The data were extracted by using Application Programmable Interface (API) from each hardware device, with the exception of Apple Watch Series 4 as a third-party software entitled "Cardiogram" was employed.

\begin{table}[tb]\setlength\tabcolsep{2pt}
\centering
\caption{Total numbers of recorded HR data point from each device.}
\label{table:datasets}
\begin{tabular}{c|cccc}
\hline
\hline
State & Fitbit (n) & E4 (n) & TicWatch (n) & Apple Watch 4 (n) \\ \hline\hline
RS & 6,063 (23) & 10,718 (24) & 16,149 (15) & 5,812 (18)\\ 
LS & 17,747 (26) & 36,207 (26) & 49,405 (16) & 17,897 (19)\\ 
IS & 11,994 (26) & 6,258 (27) & 28,217 (22) & 11,889 (28)\\ 
All & 35,804 (29) & 53,183 (29) & 93,771 (29) & 35,598 (29)\\ \hline
\hline
\end{tabular}
\caption*{ \footnotesize ``n" denotes the numbers of participant that displayed a consistent sampling rate of HR.}
\vspace{-2mm}
\end{table}

\subsection{Validation of HR estimation from wrist-worn wearable devices}
\label{sec:exp1}

\textcolor{red}{The HR$_\text{PPG}$ from each device was validated with HR$_\text{ECG}$ by using MAE.} The descriptive statistical errors in HR$_\text{PPG}$ from each device at different states are shown in Table \ref{table:DeviceErrorANOVA}. The lowest MAE in each state was emphasis with bold text. Pairwise comparison of MAE with repeated measure test is shown in Table \ref{table:pairwisecomparison}. Apple Watch Series 4 performed with significantly (p$<$0.05) lower MAE ($F(1.860,27.906) = 12.004$ in resting state, $F(1.284,19.254) = 9.448$ in \textcolor{red}{laying down} state, and $F(1.450,33.346) = 16.05$ in intense treadmill activity state) in comparison to the other devices, whereas Fitbit Charge HR yielded the highest MAE in the intense treadmill activity state. \textcolor{red}{Hence, Fitbit Charge HR, showing the highest MAE among all devices, was selected as the device of interest to investigate the improvement of HR$_\text{PPG}$ in the intense treadmill activity state}.

\begin{table}[tb]\setlength\tabcolsep{2pt}
\centering
\caption{Descriptive statistics of \textcolor{red}{the} device's HR errors with repeated measure test ANOVA.}
\label{table:DeviceErrorANOVA}
\begin{tabular}{c|cccc}
\hline\hline
\multicolumn{1}{l|}{\multirow{2}{*}{State (n)}}& \multicolumn{4}{c}{MAE $\pm$ SE (bpm)}\\ 
  & Fitbit & E4 & TicWatch & Apple Watch  \\ \hline\hline

 \multicolumn{1}{l|}{RS (13)}  & 2.95 $\pm$ 0.40  & 4.05 $\pm$ 0.34 & 2.97 $\pm$ 0.27 & \textbf{1.33 $\pm$ 0.08} \\
           \multicolumn{1}{l|}{LS (13)}  & 2.15 $\pm$  0.32  & 3.55 $\pm$ 0.34 & 2.49 $\pm$ 0.28 & \textbf{1.10 $\pm$ 0.10} \\
           \multicolumn{1}{l|}{IS (21)}  & 10.35 $\pm$  1.70 & 5.45 $\pm$ 0.48 & 4.45 $\pm$ 0.45 &\textbf{2.00 $\pm$ 0.46 }\\
           \multicolumn{1}{l|}{All (11)} & 4.22 $\pm$  1.17  & 3.90 $\pm$ 0.31 & 2.85 $\pm$ 0.22 & \textbf{1.65 $\pm$ 0.15} \\ \hline\hline
\end{tabular}
\caption*{ \footnotesize ``n" denotes the numbers of participant that displayed a consistent sampling rate of HR.}
\vspace{-1mm}
\end{table}

\begin{table}[!tb]\setlength\tabcolsep{2pt}
\centering
\caption{MAE Pairwise comparisons with repeated measure test ANOVA.}
\label{table:pairwisecomparison}
\begin{tabular}{ll|ccc}

\hline\hline
\multicolumn{2}{c|}{\multirow{2}{*}{Pair of devices}}& \multicolumn{3}{c}{MAE Mean Difference $\pm$ SE (bpm)}\\ 
 &  & RS & LS & IS  \\ \hline\hline
Fitbit & E4 & -1.02 $\pm$ 0.37& -1.24 $\pm$ 0.21 & \textbf{4.49* $\pm$ 1.46}\\
& TicWatch & 0.56 $\pm$ 0.60 & 0.39 $\pm$ 0.72    & \textbf{5.56* $\pm$ 1.44}\\
& Apple & 1.75* $\pm$ 0.37  & 1.37* $\pm$ 0.43 & \textbf{7.57* $\pm$ 1.55}  \\ \hline
E4 & Fitbit & 1.02 $\pm$ 0.37  & 1.24* $\pm$ 0.21 & -4.49$^\dagger$ $\pm$ 1.46 \\ 
& TicWatch & 1.58 $\pm$ 0.63  & 1.64* $\pm$ 0.66 & 1.07 $\pm$ 0.53            \\
& Apple & 2.77* $\pm$ 0.39  & 2.62 $\pm$ 0.40 & 3.07* $\pm$ 0.57  \\ \hline
TicWatch& Fitbit & -0.56 $\pm$ 0.60 & -0.39 $\pm$ 0.72 & -5.56$^\dagger$ $\pm$ 1.44 \\
& E4 & -1.58 $\pm$ 0.63 & -1.64$^\dagger$ $\pm$ 0.66 & -1.07 $\pm$ 0.53\\
& Apple & 1.19 $\pm$ 0.40 & 0.98 $\pm$ 0.38 & 2.01* $\pm$ 0.67  \\ \hline
Apple  & Fitbit & -1.75$^\dagger$ $\pm$ 0.37 & -1.37$^\dagger$ $\pm$ 0.43 & \textbf{-7.57$^\dagger$ $\pm$ 1.55} \\
& E4 & -2.77$^\dagger$ $\pm$ 0.39 & -2.62 $\pm$ 0.40 & \textbf{-3.07$^\dagger$ $\pm$ 0.57} \\
& TicWatch & -1.19 $\pm$ 0.40 & 0.98 $\pm$ 0.38 & \textbf{-2.01$^\dagger$ $\pm$ 0.67} 
\\ \hline\hline
\end{tabular}
\caption*{\footnotesize $*$ and $^\dagger$ indicate a significantly (at the 0.05 level) higher MAE and a significantly lower MAE, respectively.}
\vspace{-6mm}

\end{table}

\section{Result and Analysis: Heart rate post-calibration} \label{result:calibration}
The three analyses of the post-calibration of HR$_\text{PPG}$ using the ML models to improve the HR$_\text{Fitibit}$ detected by Fitbit Charge HR, the device of interest, are described and illustrated in \textit{Heart rate Post-calibration} of Figure \ref{fig:feature_analysis}: the ultimate HR$_\text{R}$ computed using selected features and ML model, model with external calculated PALs feature, and rolling regression.

\subsection{Analysis I - HR$_\text{ML}$: Improving HR estimation from the selected wrist-worn wearable device with the ML models}
Following the validation of the HR detection of the four wearable devices, the determination of the device of interest was considered depending on the highest error in HR$_\text{PPG}$. One of the four wearable devices was chosen due to its low performance. \textcolor{red}{The features from the selected device were extracted and used as the testbed in feature selection. This execution demonstrated the performance of HR estimation with the ML models, designated as HR$_\text{ML}$, using the information reported from the selected device. MAE was analyzed in all three states.} In the following set of experiments, the data from Fitbit Charge HR was chosen to be implemented with the improvement of HR$_\text{Fitbit}$ using selected features and several ML algorithms.

\label{section:exp2}


\subsubsection{Feature selection}
\label{section:feature_selection}
The features that were examined consisted of HR$_\text{PPG}$, personal health information, step count, and PAL. However, there has been no evidence that these features would benefit the model. In order to verify that each of the candidate physical features is informative and discriminative, in which the prediction of the HR$_\text{ML}$ based in these input features by our ML models essentially depends on, two basic feature selection methods were performed to test the significance of each feature to the HR$_\text{ECG}$, served as the baseline for most accurate measurement of the actual HR, in all four states (resting, \textcolor{red}{laying down}, intense treadmill activity, and all states combined). The linear relation between the feature and HR$_{\text{ECG}}$ was measured using a univariate linear regression by testing a null hypothesis that all of the regression coefficients are equal to zero. In other words, the model, which has no predictive capability using a certain feature, was tested. If the p-value of the linear relation test is smaller than 0.05, the null hypothesis can be rejected which indicates that the feature is informative and benefits the model to predict the HR$_\text{ECG}$. If the p-value is larger than 0.05, the information within the certain feature is not informative to the univariate linear regression model, which is the simplest discrimination model. In addition to the linear test, the mutual information for regression was carried out to measure the non-linear relationship, which cannot be determined by the correlation between each of the features. This relationship function is called entropy. An entropy estimated from k-nearest neighbor distances was computed following a previous study \cite{ross2014mutual} to measure the dependency between the two variables. A higher dependency level indicates a higher dependency of a certain feature with the HR$_\text{ECG}$. Battiti \textit{et al.} \cite{battiti1994using} previously investigated the dependency threshold in the mutual information test which demonstrated that the range of 0.2-0.4 was the best for trade-off point. A feature would be included in the model if it passed at least one of the two dependency tests. Means of the p-value and dependency level with SE for each state were computed across all leave-one-out cross-validation sets.
\begin{table}[t]
\centering
\caption{Univariate linear regression test and mutual information test of candidate features}
\label{table:feature_result}
\begin{tabular}{l|c@{\hspace{2em}}c@{\hspace{2em}}c@{\hspace{2em}}c@{\hspace{1em}}|c@{\hspace{2em}}c@{\hspace{2em}}c@{\hspace{2em}}c}
\hline
\hline
\multirow{2}{*}{Feature} & \multicolumn{4}{c|}{Univariate linear regression} & \multicolumn{4}{c}{Mutual information}\\
& RS & LS & IS & All& RS & LS & IS & All\\
\hline\hline
HR$_\text{\textbf{Fitbit}}$ & $*$ & $*$ & $*$ & $*$ & \checkmark & \checkmark & \checkmark & \checkmark \\
PAL &  & $*$ & $*$ & $*$ &  &  & \checkmark & \checkmark \\  
Step count & $*$ & $*$ & $*$ & $*$ & &  & \checkmark & \checkmark \\ 
Gender & $*$ & $*$ & $*$ & $*$ & \checkmark & \checkmark &  & \checkmark \\ 
PSQI & $*$ & $*$ & $*$ & $*$ & \checkmark & \checkmark &  & \checkmark \\ 
BMI& $*$ & $*$ & $*$ & $*$ & \checkmark & \checkmark & \checkmark & \checkmark \\ 
\hline\hline
\end{tabular}

\caption*{\footnotesize$*$ indicates that p-value of univariate linear test is lower than 0.05. \\\checkmark indicates that the dependency level is higher than 0.3. }
\vspace{-2mm}
\end{table}

\label{result:features_selection}
A set of features for each state was selected using univariate linear regression and mutual information tests, as reported in Table \ref{table:feature_result}. A feature was considered as an important feature if a linear relationship was found (p$<$0.05) or the dependency level between the feature and HR$_{\text{ECG}}$ is larger than 0.3. HR$_\text{\text{Fitbit}}$ and BMI displayed a dependency of the relationship with HR$_{\text{ECG}}$ in all states whereas the other features showed a distribution between the pair in at least one of the two tests. Only PAL in the resting state failed both tests and was not included in the models in the mentioned state.  
\subsubsection{ML Model selection}
\label{section:model_selection}
Six ML models (SVR, RF, GP, ANN, LR, and kNN) were trained to predict the HR$_\text{ML}$ using HR$_\text{PPG}$ from the device of interest, features from the feature selection method. Each ML model contains a kernel and different hyperparameters that required a fine-tuning to perform at its highest quality with a certain set of features. All possible sets of the hyperparameter values were compared, as a grid-search, with leave-one-out cross-validation folds. In each fold, one participant was held out as a testing set, one as a validation set, and the rest as a training set. All numerical data were normalized using standardization, based on the training set. Categorical data were encoded with zero and one before the training. The values for each hyperparameter that affect the model are displayed in Table \ref{table:hyperparameter}.

\begin{table}[t]\setlength\tabcolsep{2pt}
\centering
\caption{ A list of tuned hyperparameters \textcolor{red}{used in all models.}}
\label{table:hyperparameter}
\begin{tabular}{c|ll}
\hline
\hline
Model & Parameter/Kernel & Values\\
\hline\hline
SVR & $C$ / All & $0.001, 0.01, 0.1, 1, 10, 100$\\
&  $\epsilon$ / All & $0.001, 0.01, 0.1, 1, 10, 100$\\
&  $\gamma$ / All & $0.001, 0.01, 0.1, 1, 10, 100$\\
 & $d$ / Poly  & $2,3,4,5$\\
\hline
RF &Max features & 1, 2, 3\\
& Number of estimator & 200 - 2000 with a step of 4\\
&Max depth& 10 - 49 with a step of 3\\
&Min samples split&2 - 14 with a step of 3\\
&Min samples leaf&2 - 14 with a step of 3\\
\hline 
GP & $\alpha$ &1e-10, 1e-7, 1e-5, 1e-3, 1e-1, 1\\  
\hline
ANN & Number of hidden layer & 3, 4, 5\\
& Hidden unit in each layer
& 3 layers: (16,8,2), (16,8,4), (8,4,2)\\
& &4 layers: (16,8,4,2), (8,4,4,2)\\
& &5 layers: (16,8,4,4,2), (32,16,8,4,2)\\
& Learning rate& 0.01, 0.001, 0.0001\\
\hline
LR & $C$ & 0.001, 0.01, 0.1, 1, 10, 100\\
& Power & l1, l2, elasticnet\\
\hline
kNN & N Neighbors & 10, 20, 30, 40, 50, 100, 150, 200, \\
&&500, 1000\\
& Power & 1=Manhattan, 2=Euclidean, 3=Minkowski\\
\hline
\hline
\end{tabular}
\vspace{-4mm}
\end{table}

SVR, RF, GP, LR, and kNN models were constructed using Scikit-learn package \cite{scikit-learn}, while Keras package\cite{chollet2015keras} was used to construct a shallow ANN model.
In SVR, hyperparameters were tuned separately for each kernel. Both RBF kernel and polynomial kernel comprised $\gamma$, $C$, and $\epsilon$, in which they were tuned as core hyperparameters in SVR, where $d$ in polynomial kernel was tuned to increase the complexity of the model. All values used in the grid-search fine-tuning are shown in Table \ref{table:hyperparameter}.
Since RF has no individual kernel to be tuned, only the number of trees and early stop conditions, including the maximum number of features, the number of estimators, maximum depth, minimum sample split, and minimum sample leaf, were tuned as shown in Table \ref{table:hyperparameter}. In LR, $C$ was tuned similarly to SVR while the power function was included, whereas kNN displayed the k-nearest neighbor data points and the power hyperparameter metric that required tuning.
To construct an ANN, fully connected layer (2 and 3) with various numbers of hidden nodes were explored. The ANN was optimized using Adam optimizer with a learning ability tuned from 0.0001 to 0.1. Between each hidden layer, ReLU activation function was applied to add a non-linear transformation to the model with the exception of the last layer, which predicted the continuous value as an improved HR$_\text{ML}$.
A repeated measure test was used to test the performance of the six ML models and the hyperparameters. The MAEs of the validation sets were compared, and one ML model was adopted to be used in the subsequent experiments. The computed corrected sets of HR (HR$_\text{ML}$) from the chosen ML model were then compared with HR$_\text{ECG}$.

\label{result:model_selection}
For each of the activity state, the best performance of each ML model was tuned using the hyperparameters and validated with leave-one-out cross-validation. A set of tunable hyperparameters was selected based on the validation MAE. The results of all models are shown in Table \ref{table:tuning_result}. The result showed that SVR with RBF kernel outperformed all other models in every state with the lowest MAE. 

\begin{table}[tb]
\centering
\caption{\textcolor{red}{Mean Absolute Error of each HR estimation model evaluated in the validation set with a list of tuned hyperparameters.}}
\label{table:tuning_result}
\begin{tabular}{c|ccc}
\hline
\hline
State & Model & Hyperparameter & MAE$\pm$ SE (bpm)\\
\hline
\hline
RS & \textbf{SVR(RBF) }& \textbf{(0.001,10,1) }& $\boldsymbol{2.53\pm
0.02}$ \\
&SVR(Poly) & (3,100,0.01,0.1)&$4.89
\pm0.05$ \\
& RF &(2,800,10,2,15)&$2.90\pm0.05$ \\
& GP(RBF) & (1e-7)  & $2.64\pm0.18$\\
& ANN & (3,(16,8,4))& $6.28\pm2.65$ \\
& LR & (l1,0.1,1)& $5.06\pm0.59$ \\
& kNN & (500,2)& $4.23\pm0.527$ \\
\hline
LS& \textbf{SVR(RBF) }& \textbf{(0.001,10,0.001) }&$\boldsymbol{ 2.14
\pm0.02}$\\
&SVR(Poly) & (3,100,0.1,0.01)&$3.49\pm
0.06$\\
& RF &(2,800,10,2,15)& $2.53\pm0.02$\\
& GP(RBF) & (1e-7)   & $3.20\pm0.10$\\
& ANN & (3,(16,8,4))& $3.63 \pm 0.93$\\
& LR & (l1,0.1,1)& $4.84\pm0.53$ \\
& kNN & (100,1) & $3.35\pm0.27$ \\
\hline
IS & \textbf{SVR(RBF) }& \textbf{(0.01,10,0.1)} & $\boldsymbol{7.79\pm0.02}$\\
&SVR(Poly) & (3,0.0001,1,0.001)&$10.80\pm
0.05$\\
& RF &(2,800,10,2,15)& $8.75\pm 0.05$\\
& GP(RBF) & (1e-7) &  $19.6\pm 0.80$ \\
& ANN & (3,(16,8,4))& $11.19\pm3.01$\\
& LR & (l1,1.0)& $12.66\pm0.69$ \\
& kNN & (50,3)& $9.34 \pm0.57$ \\
\hline
All & \textbf{SVR(RBF) }& \textbf{(0.01,10,0.001)} & $\boldsymbol{4.10\pm
0.03}$\\
&SVR(Poly) & (2, 0.01,10,1)&$11.4\pm
0.07$\\
& RF &(2,800,10,2,15)&$4.57\pm0.02$ \\
& GP(RBF) & (1e-7)   &$16.68\pm0.64$ \\
& ANN & (3,(16,8,4)) &$5.32\pm0.86$ \\
& LR & (1.0,l2)& $9.42\pm0.63$ \\
& kNN & (30,1) & $5.93\pm0.57$ \\
\hline
\hline
\end{tabular}

\caption*{ \footnotesize Note: Formats of hyperparameters for each ML model are as follows: ($\gamma,c,\epsilon$) for SVR (RBF), ($d,\gamma,c,\epsilon$) for SVR (Poly), (Max features, Number of estimator, Max depth, Min samples split, Min samples leaf) for RF, ($\alpha$) for GP, (number of hidden layer, (hidden unit in each layer) for ANN, ($C$, Power) for LR, and (N Neighbors and Power) for kNN.}
\vspace{-3mm}

\end{table}
The HR$_\text{ML}$ of the testbed encompassing the PAL and the step count extracted from Fitbit Charge HR were calculated. In comparison amongst the PAL related information, the results did not provide any significant difference. Moreover, when compared to the HR$_\text{Fitbit}$ with the MAE of $3.26 \pm 0.34$ bpm in the resting state, $2.33 \pm 0.23$ bpm in the \textcolor{red}{laying down} state, $9.53 \pm 1.47$ bpm in the intense treadmill activity state, and $5.02 \pm 0.64$ bpm in all states combines. Significant improvement was present in each activity with the exception of all states combined (Table \ref{table:sensor_fusion_result}).

\subsection{Analysis II - HR$_\text{SF}$: Improving HR estimation from selected wrist-worn wearable device using sensor fusion}
\label{section:experiment2_result}
\begin{table}[b]
\centering
\caption{Univariate linear regression test and mutual information test of PAL features from fusion device.}
\label{table:fusion_feature_result}
\begin{tabular}{l|c@{\hspace{2em}}c@{\hspace{2em}}c@{\hspace{2em}}c@{\hspace{1em}}|c@{\hspace{2em}}c@{\hspace{2em}}c@{\hspace{2em}}c}
\hline
\hline
\multirow{2}{*}{Feature} & \multicolumn{4}{c|}{Univariate linear regression} & \multicolumn{4}{c}{Mutual information}\\
& RS & LS & IS & All& RS & LS & IS & All\\
\hline\hline
Crouter$_{\text{VA}}$ & $*$ & $*$ & $*$ & $*$ &  &  &  & \\  
Crouter$_{\text{VM}}$ & $*$ & $*$ & $*$ & $*$ & &  &  &  \\ 
Troiano$_{\text{VA}}$ & $*$ & $*$ & $*$ & $*$ &  & &  & \checkmark \\ 
Freedson$_{\text{VA}}$ & $*$ & $*$ & $*$ & $*$ &  &  & \checkmark & \checkmark \\ 
\hline\hline
\end{tabular}
\caption*{\footnotesize$*$ indicates that the p-value of univariate linear test is lower than 0.05. \\\checkmark indicates that dependency level is higher than 0.3.\\ VA and VM indicate that the PAL was computed based on vertical axis and vector magnitude of the accelerometer, respectively.}
\vspace{-3mm}
\end{table}

Bringing in an external sensor might be beneficial to the PALs evaluation. The PAL reported by Fitbit Charge HR was compared against the PALs calculated from the raw data from three-axis accelerometer equipped on E4, as E4 does not provide PAL of the user explicitly. Four PAL cut-points proposed by Crouter \textit{et al.} (VA and VM), Freedson \textit{et al.} (VA), and Troiano \textit{et al.} (VM) were applied to transform the raw data into PALs. VA used only y-axis in calculation whereas VM used vector magnitude. The PALs were tested with the feature selection method as described in Section \ref{section:feature_selection}. The features that passed the criteria were trained separately in each state, with the ML model and hyperparameters that yielded the lowest MAE from the previous experiment. Ultimately, a model called sensor fusion was performed to estimate HR$_\text{SF}$ with the testbed, the features from the selected device, and PAL in each state were compared among all ML models. The comparison between the PAL from the wearable device of interest and the calculated PAL from E4 was regarded as the baseline in the HR estimation process.

\label{section:result_exp3}
\textcolor{red}{The sensor fusion features for improving HR estimation models were validated in this experiment.} We used SVR with RBF kernel with a set of hyperparameters that achieved the lowest MAE from the validation set. The performance of SVR with RBF kernel on the testing set is reported in Table \ref{table:sensor_fusion_result} in comparison with other methods. The features from the testbed, i.e., PAL and step count, were enhanced by one of the four PALs from various thresholds (Crouter$_{\text{VA}}$, Crouter$_{\text{VM}}$, Troiano$_{\text{VA}}$,
wand Freedson$_{\text{VA}}$), which were introduced in Section \ref{section:PAL}. The feature selection, as in the previous experiment, was performed to test the linear relationship and the dependency between each feature and the HR$_{\text{ECG}}$. The results shown in Table \ref{table:fusion_feature_result} suggest that all PALs may possess a linear relationship with HR$_{\text{ECG}}$, although the dependencies were found only with Troino$_{\text{VA}}$ and Freedson$_{\text{VA}}$ in all states combined while only Freedson$_{\text{VA}}$ displayed dependency in intense treadmill activity state. The MAEs of the HR$_\text{SF}$ derived from the PALs of E4 using the method of sensor fusion are reported in Table \ref{table:sensor_fusion_result}.

\begin{figure}[tb]
\centering
\begin{subfigure}{.40\textwidth}
\includegraphics[width=0.85\textwidth]{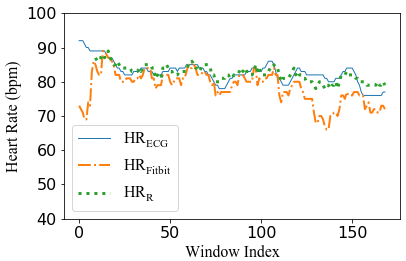}\vspace{2mm}
 \caption{Resting Stage}
  \label{fig:timeseries-resting}
\end{subfigure}

\begin{subfigure}{.40\textwidth}
\includegraphics[width=0.88\textwidth]{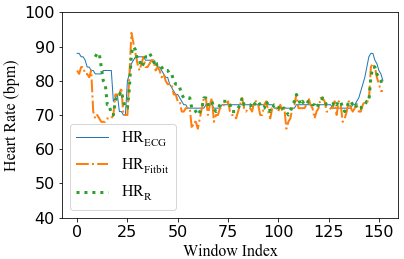}\vspace{2mm}
 \caption{\textcolor{red}{Laying down} Stage}
  \label{fig:timeseries-sleeping}
\end{subfigure}

\begin{subfigure}{.40\textwidth}
\includegraphics[width=0.88\textwidth]{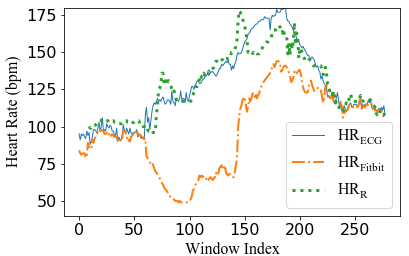}\vspace{2mm}
 \caption{Intense Treadmill Activity Stage }\vspace{2mm}
  \label{fig:timeseries-intensity}
\end{subfigure}

\caption{Estimation results from Participant 11. A comparison between the ground truth of HR (HR$_\text{ECG}$) and HR computed by Fitbit Charge HR (HR$_\text{Fitbit}$) according to the three activity states; resting, \textcolor{red}{laying down}, and intense treadmill activity states. The window indexes are from HR$_\text{Fitbit}$ every 15 seconds.}
\label{fig:timeseries_plot}
\vspace{-4mm}
\end{figure}

\subsection{Analysis III - HR$_\text{R}$: Improving HR estimation from selected wrist-worn wearable device using rolling regression}
\label{section:exp4}
Improving the HR$_\text{ML}$ using the ML model with only PAL and personal health information does not fully utilize the ML algorithms and data as only a few features (low dimensionality) were trained to make a prediction in the previous experiment. Furthermore, it does not exploit the temporal information from the input, which is a time-series. To address these problems, a rolling regression was adopted, extended from the previous experiment, by considering a few data points in the time domain instead of only a single data point. A window of features was built using device-estimated HR$_\text{PPG}$, PAL, and step count within a window size, recorded every 15 seconds, controlling the amount of sequential data points to be utilized (5, 10, and 15 data points). Although the drawback of the rolling regression, in which a first few data points at the beginning in the time domain could not be predicted, is unavoidable, it was used as features to predict the subsequent HR$_\text{R}$. To compensate for the drawback, the chosen amounts of window size were optimized to be the smallest sizes that would not be considerably underperformed by the larger window size. For instance, a window size of 20 data points would affect the protocol as it would require up to 5 minutes of the data points. Therefore, the window sizes of 5, 10, and 15 data points were nominated for the comparison to determine the best possible window size. Similar to the previous experiment, the selected features, including gender, PSQI, and BMI, were used as the personal health information features without the rolling window. The performance of the selected rolling window model of the rolling regression was compared to all previous experiments using MAE: HR$_\text{ML}$, HR$_\text{SF}$, and HR$_\text{R}$.

\label{section:result_exp4}
Using the method of rolling regression, the ML algorithms exploited the features over the temporal domains to create a prediction. In this study, the temporal features were obtained from Fitbit Charge HR, including the HR$_\text{Fitbit}$, PAL, and step count. The sizes of the rolling window were tuned to match 5, 10, and 15 data points. The results of the three window sizes are reported in Table \ref{table:sensor_fusion_result} (Analysis III). The results between the HR$_\text{PPG}$ obtained from the devices, the HR$_\text{R}$, and the baseline (HR$_\text{ECG}$) were compared in Figure \ref{fig:timeseries_plot} in time-series.

\begin{figure}[]
\includegraphics[width=72mm]{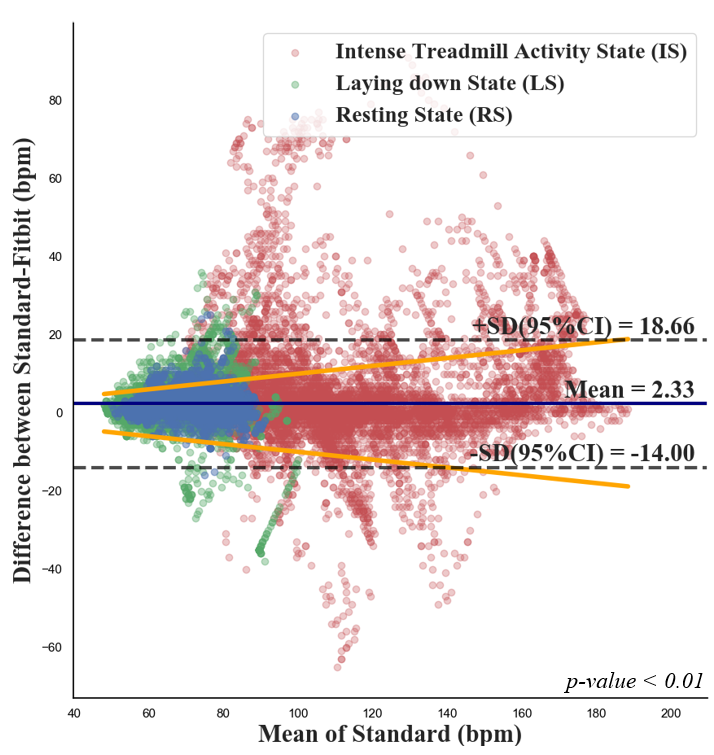}\vspace{2mm}
  \caption{Bland-Altman plot  displaying the difference in HR between the standard measurement (HR$_\text{ECG}$) and Fitbit Charge HR (HR$_\text{Fitbit}$) before post-calibration with the rolling regression model (HR$_\text{R}$). The yellow line indicates an acceptable interval.}
  \label{fig:blandaltman-before-correction}
  \vspace{-5mm}

\end{figure}
\begin{figure}[]
\includegraphics[width=72mm]{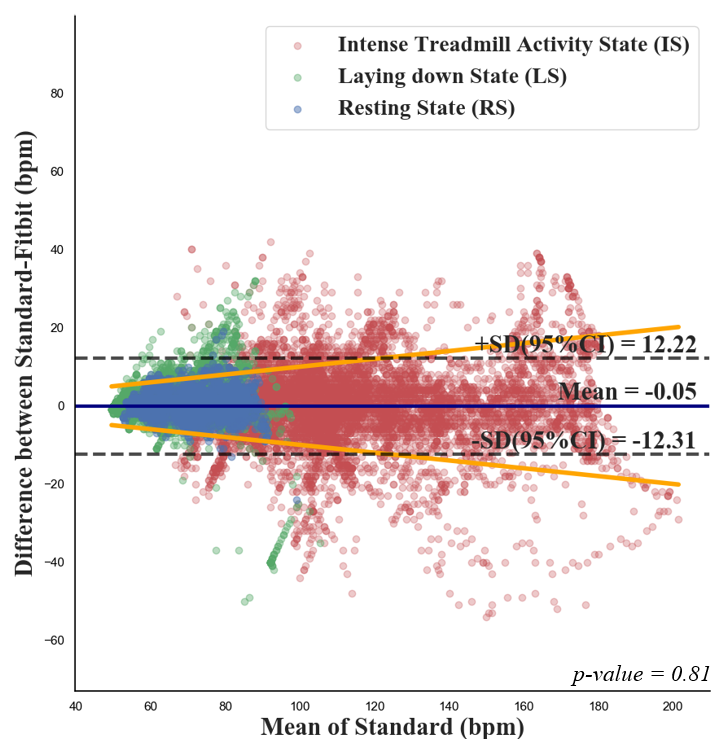}\vspace{2mm}
  \caption{Bland-Altman plot displaying the difference HR between the standard measurement (HR$_\text{ECG}$) and Fitbit Charge HR (HR$_\text{Fitbit}$) after post-calibration with the rolling regression model (HR$_\text{R}$). The yellow line indicates an acceptable interval.}
  \label{fig:blandaltman-after-correction}
  \vspace{-3mm}
\end{figure}
Bland-Altmans plots were used to compare the HR$_\text{ECG}$ and the HR$_\text{Fitbit}$ reported by Fitbit Charge HR. The confidence interval was set at 95\%. Before the correction, Figure \ref{fig:blandaltman-before-correction} shows 1,052 points out of the confidence interval range while Figure \ref{fig:blandaltman-after-correction} shows 406 points after correction. The two-tailed t-test between HR$_\text{Fitbit}$ and the correction of HR$_\text{R}$ using the rolling window regression with the window size of 10 indicated the error reduction of 33.44\% ($t(5323)=16.21$, p$<$0.05) in the resting state, 15.88\% ($t(15013)=9.82$, p$<$0.05) in the \textcolor{red}{laying down} state, 9.55\% ($t(8477)=7.65$, p$<$0.05) in the intensity state, and 18.73\% ($t(19313)=19.71$, p$<$0.05) in all states combined.


\begin{table*}[bt]\setlength\tabcolsep{2pt}
\centering
\caption{MAE of the HR estimation using the post-calibration approach with PAL related information from Fitbit Charge HR, sensor fusion PAL, and rolling regression of information from Fitbit Charge HR as main features on the testing set.}
\label{table:sensor_fusion_result}
\begin{tabular}{l>{\centering}p{3.5cm}>{\centering\arraybackslash}p{3.5cm}>{\centering\arraybackslash}p{3.5cm}>{\centering\arraybackslash}p{3.5cm}}
\hline
\hline
\multirow{2}{*}{HR estimation} & \multicolumn{4}{c}{MAE$\pm$ SE (bpm)} \\

& RS & LS & IS & All \\
\hline\hline\\[-1.5ex]
\multicolumn{5}{l}{(Heart Rate Validation) \textit{HR$_\text{Fitbit}$ from Fitbit Charge HR.}}\\
\hspace{1em}Fitbit & $3.26 \pm 0.34$ & $2.33 \pm 0.23$ & $9.53 \pm 1.47$ & $5.02 \pm 0.64$ \\ \hline \hline

\multicolumn{5}{l}{Heart Rate Post-Calibration}\\
\multicolumn{5}{l}{\textit{(Analysis I -  HR$_\text{ML}$) Using improving HR$_\text{Fitbit}$ estimation model with PAL related information from Fitbit Charge HR. }}\\
\hspace{1em}PAL & *$2.90 \pm 0.35$ & *$2.15 \pm 0.20$ & *$9.25 \pm 1.32$ & $5.10 \pm 0.65$ \\

\hspace{1em}Step count & *$2.92 \pm 0.34$ & *$2.16 \pm 0.20$ & *$9.24 \pm 1.40$ & $4.92 \pm 0.54$ \\

\hspace{1em}PAL and step count& \multirow{1}{*}{*$2.92 \pm 0.34$} & \multirow{1}{*}{*$2.16 \pm 0.20$} & \multirow{1}{*}{*$9.15 \pm 1.31$} & \multirow{1}{*}{$4.88 \pm 0.54$}  \\

\hline\\[-1.5ex]
\multicolumn{5}{l}{\textit{(Analysis II -  HR$_\text{SF}$) Using improving HR$_\text{Fitbit}$ estimation model with sensor fusion PALs as the main feature. }}\\
\hspace{1em}Crouter$_{\text{VA}}$ & *$2.86 \pm 0.34$ & *$2.20 \pm 0.19$ & *$9.47 \pm 1.41$ & $5.02 \pm 0.69$ \\

\hspace{1em}Crouter$_{\text{VM}}$ & *$2.87 \pm 0.33$ & *$2.17 \pm 0.18$ & *$9.41 \pm 1.42$ & $5.02 \pm 0.69$ \\

\hspace{1em}Troiano$_{\text{VA}}$ & *$2.86 \pm 0.03$ & *$2.24 \pm 0.20 $ & $ 9.60 \pm 1.43$ & $5.12 \pm 0.72$ \\

\hspace{1em}Freedson$_{\text{VA}}$ & *$2.87 \pm 0.34$ & *$2.23 \pm 0.20$ & *$9.68 \pm 1.44$ & $5.14 \pm 0.72$ \\[1px]

\hline\\[-1.5ex]

\multicolumn{5}{l}{\textit{ (Analysis III -  HR$_\text{R}$) Using improving HR$_\text{Fitbit}$ estimation model with a rolling window and PAL related information from Fitbit Charge HR as features. }}\\
\hspace{1em}5 points & *$2.55 \pm 0.40$ & *$1.99 \pm 0.22$ & *$8.65 \pm 1.05$ &  *$4.15 \pm 0.34$\\[1px]
\hspace{1em}10 points & *$2.17 \pm 0.18$ & *$1.96 \pm 0.22$ & *$8.62 \pm 1.04$ & *$4.08 \pm 0.32$\\[1px]
\hspace{1em}15 points & *$2.16 \pm 0.18$ & *$1.97 \pm 0.23$ & *$8.60 \pm 1.03$ & *$4.03 \pm 0.30$
\\[0px]
\hline
\hline
\end{tabular}
\caption*{\footnotesize*An error reduction is significant when compared with HR$_\text{Fitbit}$ (p$<$0.05).}
\vspace{-5mm}
\end{table*}

\section{Discussion}
\label{Discussion}

The accuracy of the PAL assessment performed by the wrist-worn fitness trackers is imperative to the monitoring of the fitness of physical conditions and health. Numerous, yet validation voids, wrist-worn portable devices have been attractively targeting health-concerned purchasers in the rising wearable product market. Here in this study, we seek to evaluate the validity of the four chosen devices including Fitbit Charge HR, Apple Watch Series 4, TicWatch Pro, and E4 as well as the methods to improve the HR estimation during post-calibration using the data provided by these devices. One of the primary aims is to establish a universal experimental protocol including the three states of physical activity which mimic the daily activities in real life: resting, \textcolor{red}{laying down}, and intense treadmill activity states. Several studies have exploited different sets of study protocols with similar protocols involving active exercising on a treadmill and a cycle ergometer \cite{wallen2016accuracy, bai2018comparative, tedesco2019accuracy}. Although the studies differentiated between the activities at rest (sitting and lying), low intensity activities (household chores), and high intensity activities (walking and running), to our best knowledge, the protocol used in the present study entails different levels of active intensity as well as the \textcolor{red}{laying down} state. This may give rise to further insights on the HR measurement in the daily-like environment.  
The results from the validation of the four chosen devices demonstrated and confirmed that the MAEs of the HR$_{\text{PPG}}$ from all the devices measured in the intense treadmill activity state exhibited higher than the other two states. The high MAE was in line with our initial hypothesis that MA might have caused an inaccuracy in the measurement by the PPG sensors. The comparison manifested that Fitbit Charge HR performed with a significantly higher MAE than TicWatch Pro and Apple Watch Series 4, and higher, but no significance, than E4 during the intense treadmill activity state. Interestingly, E4, the only medical grade class 2a wearable device from Empatica, performed with the second-highest MAE in all states, followed by TicWatch Pro. Despite the interference by MA, Apple Watch Series 4 achieved with the lowest MAE in all states compared to the other three devices. During the data collection in intense treadmill activity state, HR$_{\text{PPG}}$ from E4 could not be collected at a proper rate due to the HR estimation algorithm on the apparatus that disregards the HR$_{\text{PPG}}$ if peak-to-peak in the PPG signal is not obviously detected. To comprehensively develop the model for improving the HR estimation during the post-calibration process, only the HR$_{\text{Fitbit}}$ was considered as an input of the model, based on the pairwise comparison with repeated measure test of the MAE achieved by each pair of the devices. Fitbit Charge HR \textcolor{red}{generated} significantly higher MAE in the intense treadmill activity state which severely affected by MA.

It could be speculated that due to the lower amount of hardware components equipped on Fitbit Charge HR, which comprises one PPG sensor, two green LEDs, 1 Photodiode, three-axis accelerometer, and without a gyroscope, the accuracy of the HR$_{\text{Fitbit}}$ may perhaps be affected. However, the accuracy produced by each device may depend on the in-house algorithm. As displayed in Table \ref{table:device-compare}, although TicWatch Pro contains similar components to Fitbit Charge HR with an additional gyroscope, the device performed with a lower error rate than the medical grade E4. Thus, this suggests the importance of MA removal methods.

One appealing aspect of Fitbit Charge HR encompasses its functions of PAL computation, step count, and HR estimation simultaneously, in which we presumed that using these recorded data as the model features can improve the HR estimation during the post-calibration process, particularly in the intense treadmill activity state where the MA is likely to occur. From the MAE of HR$_\text{ML}$ in Section \ref{section:experiment2_result}, SVR with RBF kernel outperformed the other ML algorithms, i.e., SVR with polynomial kernel, RF, GP, ANN, LR, and kNN, in all states. The error in HR$_\text{Fitbit}$ on Fitbit Charge HR, which was severely affected by MA, could be lower using the ML model with the features that are regularly recorded by the device. The PAL and step count from Fitbit Charge HR were also compared. The models using PAL with step count as features slightly performed better in the intense treadmill activity states and all states, note that the models with separated features (Table \ref{table:sensor_fusion_result}) exhibited no difference. However, it is still not conclusive whether the PAL and step count solely would \textcolor{red}{mainly} benefit the HR estimation.

Due to the raw three-axis accelerometer data reported by E4, the four most adopted PAL estimation methods proposed by Freedson \textit{et al.} \cite{pamty2005calibration}, Troiano \textit{et al.} \cite{troiano2008physical}, and Crouter \textit{et al.} \cite{crouter2015estimating} were used to calculate the accelerometer-based PALs as a feature, as displayed in Table \ref{table:sensor_fusion_result}. The MAEs on Crouter$_{\text{VA}}$, Crouter$_{\text{VM}}$, and Freedson$_{\text{VA}}$ were significantly reduced in each activity except in all states whereas Troiano$_{\text{VA}}$ only displayed significant reduction in resting and \textcolor{red}{laying down} states. Furthermore, while the data from the three-axis accelerometer in Fitbit Charge HR could not be applied in this study, E4 contains additional axes of both VA and VM. However, the testbed containing the PAL and the step count acquired from Fitbit Charge HR was found to be similar to the four PALs calculated from E4 with no significant value exhibited. Thus, this method did not demonstrate that the PAL feature obtained from the external three-axis accelerometer could noticeably improve the calibration model when compared to the testbed.

The model using the PAL, step count, and rolling window sizes of 5, 10, and 15 points revealed a significant error reduction from HR$_\text{Fitbit}$ to HR$_\text{R}$ (p$<$0.05) in all activity states, especially in all states in which the error reduction was shown in the previous analysis using sensor fusion (Analysis II) as non-significant (p$>$0.05). In comparison between the three window sizes, no significant difference was achieved. Nonetheless, the rolling regression exhibited further significant improvement beyond the other methods on the testing set in the previous analyses. Although all sizes of the rolling window also displayed an improving trend of HR estimation, the rolling window size of 15 showed the most significance \textcolor{red}{compared to the other window sizes of 5 and 10 regarding the testbed.} It is reasonable to select the rolling window size of 10 as the finest performance of all the methods analyzed. Firstly, it performed with lower errors when compared to 5 points. Secondly, although 15 points might have performed with less errors when compared to 10 points in some states, it could compensate for the drawback of disregarding several more data points while achieving the significant improvement to reduce the error in which the window size of 15 would have. Interestingly, \textcolor{red}{the HR improvement was found in the laying down state, which was expected to create a lower production of MA errors as the human body would generally not create a fair amount of} movement during sedentary activity \cite{ram2011novel}. Furthermore, a significant difference was detected in the intense treadmill activity state, which is the state of interest to reduce the MA and improve the HR estimation using the post-calibration approach. This suggests a further improvement of the proposed methods for future investigation.

The public dataset produced by TROIKA is widely used and \textcolor{red}{has} been tested using different estimation models such as TROIKA, WPFV, and CorNET. However, these models did not engage in the post-calibration process. The works were applied with raw PPG signals to remove MA. Therefore, we deem that these frameworks are not comparable to our proposed method.

The post-calibration approach has \textcolor{red}{the} potential to be established as a standard practice to further improve the accuracy of the HR estimation both in the consumer grade and medical grade wearable products. However, many users have expressed concerns over the security mechanisms within the wearable devices as the personal and medical information \textcolor{red}{is} stored and collected. Recently, Zheng and colleagues have analyzed different cryptographic primitive techniques to facilitate and distribute the data from the sensors of the portable medical ECG devices in order to protect the wireless devices securely \cite{zheng2018critical}. Furthermore, the security within the wireless body sensor networks (WBSNs) has been explored to improve the heartbeats and the \textcolor{red}{signal} processing time, including noise removal, encoding technique, and feature extraction \cite{pirbhulal2018heartbeats}. Thus, future studies are encouraged to investigate the security system during the data collection for the post-calibration approach while draw a definitive framework to improve the efficacy of the post-calibration methods in various wearable devices.

\vspace{-2.2mm}\section{Conclusion}
\label{Conclusion}
This study presents the investigation on the accuracy of HR estimation in the four popular portable wrist-worn wearable devices: Fitbit Charge HR, Apple Watch Series 4, TicWatch Pro, and E4. Although these devices are favorable due to its convenience, the PPG signals affected by MA are not accurate. The devices are valid and reliable for measuring HR in sedentary and low PAL, whereas a higher error rate is detected during moderate to vigorous PAL. The HR correction should be performed before the report, especially in Fitbit Charge HR. A few post-calibration approaches to improve the HR estimation from PPG sensors are proposed in this study. Ultimately, the model with the rolling window was verified to yield the corrected HR with the least error. 
Furthermore, the benefits of this validation would undoubtedly encourage the manufacturers of wearable devices to adopt these approaches towards consumers. The post-calibration would be propitious to the non-real-time analysis of \textcolor{red}{the} HR dataset, collected from the wearable devices, which may consequentially aid the calculation of energy expenditure and the diagnosis of CVD. 

\ifCLASSOPTIONcaptionsoff
  \newpage
\fi

\vspace{-2.2mm}\bibliographystyle{IEEEtran}
\bibliography{ref} 
\end{document}